# Encoding of direct 4D printing of isotropic single-material system for double-curvature and multimodal morphing

Bihui Zou[1], Chao Song[1], Zipeng He[2], Jaehyung Ju[1]

[1] UM-SJTU Joint Institute, Shanghai Jiao Tong University

[2] School of Aeronautic and Astronautic Engineering, Shanghai Jiao Tong University

800 Dongchuan Road, Shanghai 200240, China

**Abstract**:

The ability to morph flat sheets into complex 3D shapes is extremely useful for fast manufacturing and saving materials while also allowing volumetrically efficient storage and shipment and a functional use. Direct 4D printing is a compelling method to morph complex 3D shapes out of as-printed 2D plates. However, most direct 4D printing methods require multi-material systems involving costly machines. Moreover, most works have used an open-cell design for shape shifting by encoding a collection of 1D rib deformations, which cannot remain structurally stable. Here, we demonstrate the direct 4D printing of an isotropic single-material system to morph 2D continuous bilayer plates into doubly curved and multimodal 3D complex shapes whose geometry can also be locked after deployment. We develop an inverse-design algorithm that integrates extrusion-based 3D printing of a single-material system to directly morph a raw printed sheet into complex 3D geometries such as a doubly curved surface with shape locking. Furthermore, our inverse-design tool encodes the localized shape-memory anisotropy during the process, providing the processing conditions for a target 3D morphed geometry. Our approach could be used for conventional extrusion-based 3D printing for various applications including biomedical devices, deployable structures, smart textiles, and pop-up Kirigami structures.

**Significance**:

Most 4D printings utilize 1D structural deformation — bending and coiling to morph a flat shape by controlling only a single curvature, limiting the realization of complex 3D curved surfaces with two curvatures. We quantitively analyze the morphing of a 2D plate into doubly curved surfaces on a direct 4D printing of a single isotropic material employing anisotropy of shape memory effect during the extrusion-based printing. We also construct an inverse-design algorithm to identify the printing conditions for a target 3D shape, such as the extruding direction per layer, the thickness ratio of the layers, the melting temperature, and the activation temperature.

# I. Introduction

Morphing flat sheets into complex 3D geometries is significant from engineering, economic, and artistic perspectives, as manufacturing a product in the form of 2D plates before deployment can facilitate storage and transportation [1,2]. 4D printing technologies can enable 3D morphing from as-printed flat sheets [1-5] when exposed to environmental stimuli such as heat [6-9], moisture [3,10], and light [11,12]. Unlike conventional 4D printing requiring additional programming steps for deployment after 3D printing, direct 4D printing integrates the programming with a printing procedure [3,4] that can directly transform a 2D plate into a 3D curved shape. Embedding this programming strategy into the printing procedure is significant because the as-printed 2D plate does not require supporting materials, saving material during the printing procedure [3-5]. Moreover, morphing as-printed 2D plates has an advantage over printing 3D shapes because of the tremendous savings in volume for both storage and transportation [5]. Despite its vast significance, there are few works on direct 4D printing because of the challenge of fully integrating the printing process with simultaneous morphing to complex 3D shapes such as doubly curved surfaces.

To date, there are two general categories for direct 4D printing according to their triggering mechanisms. One utilizes the anisotropy of printing filaments, e.g., the direct ink writing (DIW) of a hydrogel with fibrils that can produce anisotropic swelling strain when submerged in a solvent [3]. The other employs a stress mismatch of two materials [4,5,13,14]; e.g., a volumetric shrinkage mismatch between elastomer and hydrogel layers during dehydration [13] and a swelling mismatch of a composite with hydrophilic PEGDA and hydrophobic PPGDMA [14]. Despite its excellent resilient properties, the former has several drawbacks — low stiffness, slow response of swelling, and unstable actuated shape [3]. The latter involves a complicated synthesis of two materials and requires supporting materials [4,5].

Moreover, the current level of direct 4D printing uses 1D structural deformation — bending and coiling to morph a flat shape by controlling only a single curvature along the longitudinal direction. Although many works demonstrated 1D rod[5], 2D plate[15], or even 3D tube [16] to transform into 3D structures, their principle utilizes a 1D bending deformation. However, in nature, there are continuous 3D curved surfaces described by two curvatures mutually perpendicular to each other or Gaussian curvature (multiplication of two principal curvatures), e.g., a saddle shape where positive and negative curvatures can describe the geometry in two directions. One group has demonstrated 4D printing of a doubly curved surface with a porous structure by discretizing the surface into multiple units [2]. However, their deployment principle still uses a 1D deformation of lattices' ribs. The porous structures may not suit general engineering applications such as waterproofing, sealing, and continuous structures. Notably, direct 4D printing from a continuous 2D flat surface into a complex curved 3D shape such as a doubly curved surface with negative Gaussian curvature is exceptionally challenging due to the coupling of bending and twisting during the deployment. Besides, most previous works are a qualitative study of deployment [15,16] or a simple quantitative analysis of 1D bending and coiling deformations [5,6,13].

To overcome the technical challenges of direct 4D printing, including multi-material printing, a complicated synthesis procedure of anisotropic printing inks, and limited 1D deployment[1-5], we quantitively analyze the morphing of a 2D plate into doubly curved and multimodal 3D geometries on a direct 4D printing of a single isotropic material. Moreover, one can lock a morphed shape after deployment while the local stress releases. To better understand the processing–morphing relationship, we combine laminated plate theory with the differential geometry of Gaussian curvature, verifying the model with experiments to directly morph as-printed sheets into complex 3D geometries. We also construct an inverse-design algorithm to identify the printing conditions for a target 3D shape, such as the extruding direction per layer, the thickness ratio of the layers, the melting temperature, and the activation temperature. By applying the extrusion-based 3D printing of an isotropic shape memory polymer (SMP) to the direction-dependent extrusion for each layer, we reveal four major transformation modes that can produce a complex freeform surface in 3D space: in-plane stretching/shear and out-of-plane bending/twisting, where we decompose the modes for the construction of the inverse-design algorithm. Our direct 4D printing method of a single-material system can be applied to conventional extrusion-based 3D printing such as Fused Deposition Modeling (FDM) and Direct Ink Writing (DIW) for various morphing structures with shape locking [17], biomedical devices [18-20], electronics [21,22], smart textiles [23], and Origami/Kirigami structures [24-26].

## II. Overall concept

Examining the extrusion-based 3D printing of polymers at an elevated temperature above the glass-transition temperature $T_g$, we find a strong analogy between the printing process and thermomechanical training of SMPs, as illustrated in Figure 1a. For example, during FDM printing, a typical extrusion-based 3D printing method, one can move a thermoplastic filament at room-temperature $T_L$ (stage 1) into a nozzle at high temperature $T_H(>T_g)$ (stage 2). The transferred filament is squeezed into a nozzle at $T_H$ for printing, analogous to the mechanical programming of the conventional SMP at $T_H$, as illustrated in stage 3 in Figure 1a. After printing, the extruded filament cools down, solidifying while retaining significant internal stress along the printing direction and minor stress by bonding with adjacent filaments perpendicular to the printing direction, as illustrated in stage 4 in Figure 1a. If one reheats the printed part to a temperature above $T_g$ of the material, the part recovers to a stress-free condition at a low-energy state, as shown in stage 5 in Figure 1a. Specifically, the printed part shrinks along the printing direction and expands in the transverse direction due to Poisson's effect, as illustrated in Figure 1b. Because of the large intrinsic anisotropic flow by extrusion during the printing procedure above $T_H$, we can produce sizeable anisotropic deformation of printed parts during the recovery of the shape memory effect. Moreover, varying printing directions for each layer can produce both inter-lamina and intra-laminar stresses by coupling and decoupling the directional thermal strain mismatch during the printing procedure. Notably, a bilayer plate has a different directional shape-memory recovery between the top and bottom layers, producing curvature due to the forced contact along the entire midplane. Eventually, the direct 4D printing can produce four modes of thermomechanical transformation of laminates: two in-plane modes, axial and shear, and two out-of-plane modes, bending and twisting, as shown in Figure 1c. Therefore, the embedded anisotropic thermomechanical training of the SMP during the printing procedure causes an anisotropic recovery of filaments coupled with deformation of a laminate, enabling deployment of the as-printed structures with multimodal and complex 3D shapes. Eventually, exploring the mechanics of the printing process and deployment can help us develop an inverse-design algorithm.

This intrinsic thermomechanical training of a SMP during extrusion-based printing has been previously identified [15,27]. However, the relevant previous works did not construct an analytical model capable of tackling the forward- and inverse-design problems; only qualitative works are available to demonstrate simple 1D structural deformations, e.g., only the bending and coiling modes of beams [15,27]. In the current work, we analytically encode the direct 4D printing of 2D plates to complex freeform surfaces, e.g., the wrinkled circular plate in Figure 1d and doubly curved plates, which can be validated with experiments and finite-element (FE) simulations. Previously, generating a wrinkled pattern such as that in Figure 1d required planar buckling due to the confinement of flat sheets [28]. However, our method can generate a wrinkled shape without compressive buckling due to an in-plane shrinkage combined with out-of-plane bending/twisting during the deployment (see Supplementary Video 1).

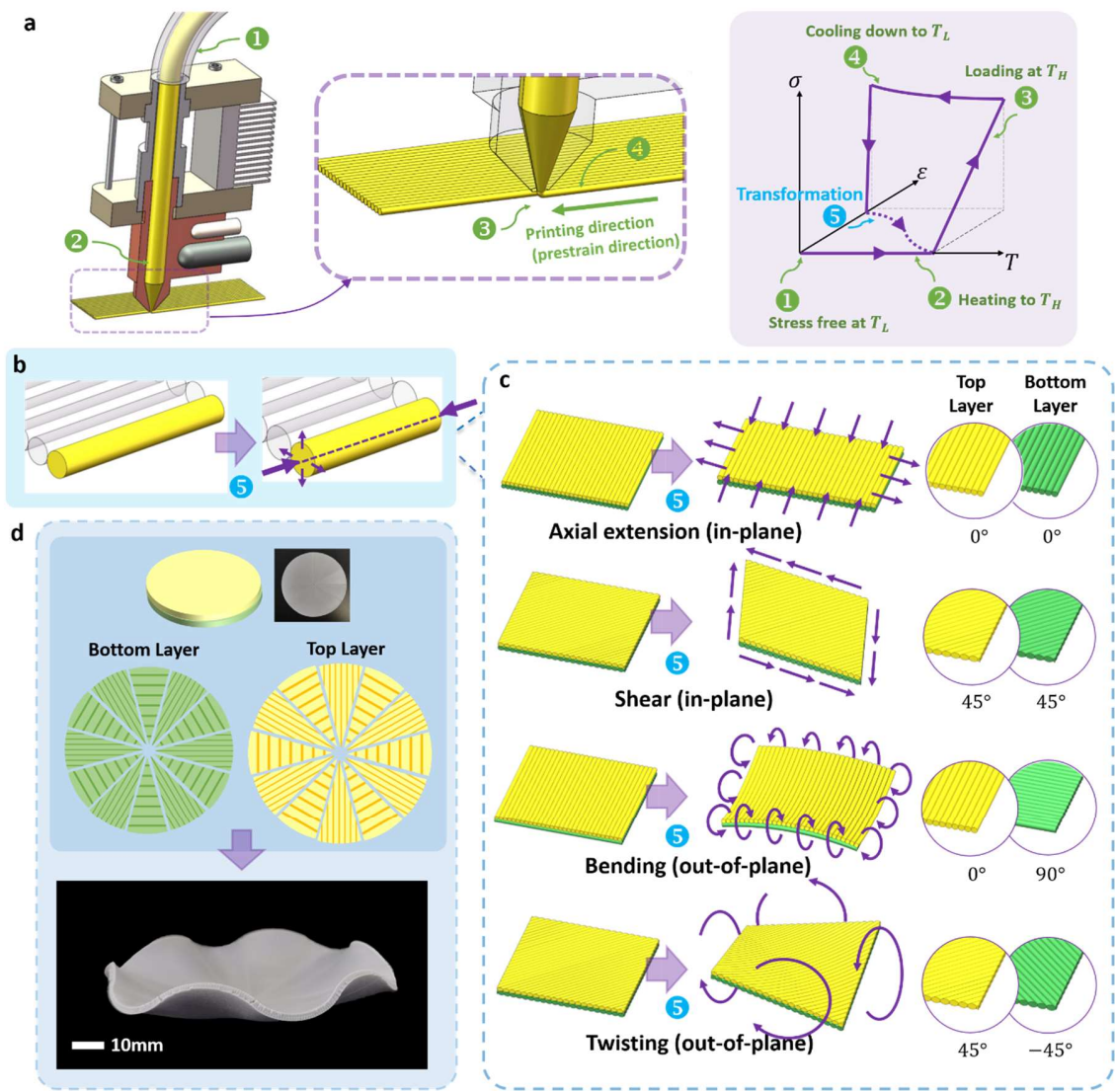

*Figure 1. Principle of direct 4D printing of a single material: (a) analogy of printing process with anisotropic thermomechanical training of SMP filaments, (b) anisotropic recovery of SMP filaments during deployment, (c) four transformation modes of the direct 4D printing of 2D bilayer plates, and (d) example of direct 4D printing — a wrinkled bilayer circular plate.*

## III. Integrated modeling of multimodal and double-curvature deformations

We integrate the printing process, geometry, and materials, providing multimodal transformation of a bilayer plate. Using the laminated plate theory [29-31] combined with the 3D printing process, we develop a theoretical model for in-plane and out-of-plane deformation of a bilayer ($n = 2$) plate during the thermomechanical deployment:

$$\mathbf{N} = \mathbf{A} \cdot \boldsymbol{\varepsilon} + \mathbf{B} \cdot \boldsymbol{\kappa} - \mathbf{N}^T$$
$$\mathbf{M} = \mathbf{B} \cdot \boldsymbol{\varepsilon} + \mathbf{D} \cdot \boldsymbol{\kappa} - \mathbf{M}^T, \qquad (1)$$

where $\mathbf{N} \left(= \sum_{k=1}^{n} \int_{z_{k-1}}^{z_k} \boldsymbol{\sigma}\, dz\right)$ and $\mathbf{M} \left(= \sum_{k=1}^{n} \int_{z_{k-1}}^{z_k} \boldsymbol{\sigma} z\, dz\right)$ are the resultant force and moment vectors, respectively, for the in-plane stress $\boldsymbol{\sigma}\,(\sigma_x, \sigma_y, \sigma_{xy})$ on a lamina. $\mathbf{A}$, $\mathbf{B}$, and $\mathbf{D}$ are the extension, extension–bending coupling, and bending tensors, respectively. $\mathbf{N}^T$ and $\mathbf{M}^T$ are the resultant thermal force and moment vectors by the extruded filament during 3D printing. Eventually, Equation (1) provides the in-plane strain $\boldsymbol{\varepsilon}$ and out-of-plane curvature $\boldsymbol{\kappa}$ in vector form, providing a deformed shape after deployment. Note that a bilayer is a minimum necessary condition to generate all four deformation modes in Figure 1c. A more detailed description of Equation (1) is provided in Section S1 of the Supplementary Information.

Note that the thermomechanical deployment in Equation (1) is integrated with the recovery behavior of polylactic acid (PLA), a typical material of an FDM 3D printing. The dynamic mechanical analysis (DMA) results in Figure 2a show the phase transition of PLA from a glassy to rubbery state with a vast stiffness drop near its $T_g$ of 60.2°C, implying the minimum activation temperature $T_a$ for deployment. For varying $T_a$, we measured the longitudinal shrinkage $l/l_0$ of PLA as a function of exposure time, as shown in Figure S7 in the Supplementary Information, where we can estimate the time of a full deployment for varying $T_a$; e.g., $t = {\sim}180\ s$ for $T_a = 85°C$. A higher $T_a$ indicates more significant longitudinal shrinkage and greater lateral expansion [32], as shown in Figure 2b. Note that we measure the deployment behavior for a specific printing condition: a nozzle traveling speed of $60\ mm/s$, a nozzle temperature of $210°C$, a layer height of $0.1\ mm$, and a flow compensation of $100\%$. We can estimate the thermal strains $\varepsilon_1$ and $\varepsilon_2$ of PLA for varying printing conditions, as shown in Figure 2c. Previously, a nozzle speed was used to control the anisotropy of a composite filament in DIW with viscous shear flow [2]. However, the nozzle speed does not affect the anisotropy of our filament, as shown in Figure 2d. Instead, the temperature is a more dominant factor in controlling the anisotropy because the main principle to induce anisotropy is a shape memory effect, as confirmed in Figure 2c.

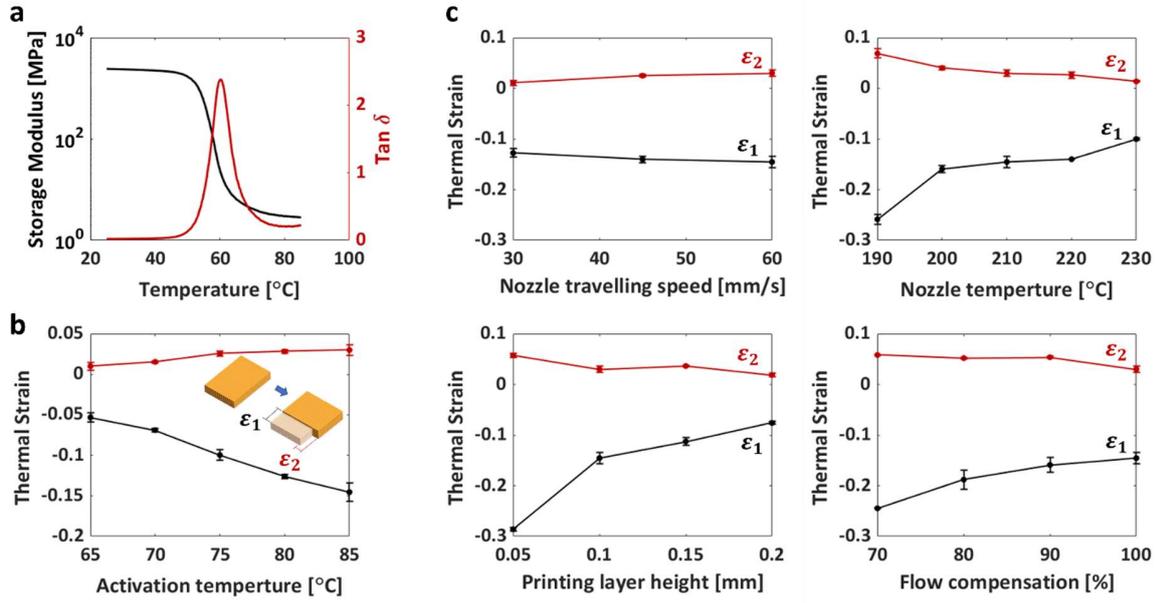

*Figure 2. Characterization of PLA and printed single-layers: (a) storage modulus and $\tan \delta$ of PLA, (b) steady-state thermal strains of printed single-layers for varying $T_a(> T_g)$ measured at $t_a = 180\ s$, (c) thermal strains of single layers at $T_a = 85°C$ and $t_a = 180\ s$ for varying process parameters — nozzle traveling speed, nozzle temperature, printed layer height, and flow compensation.*

Figure 3 shows the in-plane and out-of-plane deformations for varying printing angles ($\theta_1$ and $\theta_2$) and thickness ratios for $T_a = 85°C$. The in-plane deformation modes appear when three curvatures are zero ($\kappa_x = \kappa_y = \kappa_{xy} = 0$). We can generate in-plane shrinkage ($\gamma_{xy} = 0$) at the center and corners of the domain of $-90° \leq \theta_1, \theta_2 \leq 90°$, as shown in Figure 3a. Note that we only have two cases for the pure in-plane shrinkage: $[0°, 0°]$ and $[90°, 90°]$; $[-90°, -90°]$, $[90°, -90°]$, $[90°, 90°]$, and $[-90°, 90°]$ are all geometrically equivalent. In contrast, in-plane shear deformation ($\gamma_{xy} \neq 0$) lies in the solid purple diagonal line of the domain, as shown in Figure 3, where the printing directions of the top and bottom layers are the same except for two aligning cases: $[0°, 0°]$ and $[90°, 90°]$. Notably, $[0°, 0°]$ and $[90°, 90°]$ provide the maximum shrinkage during the deployment in the $x$-and $y$-directions, respectively. The maximum in-plane shear deformation occurs at $[45°, 45°]$ and $[-45°, -45°]$, as shown in Figure 3a.

Mismatching the printing directions of each layer can produce out-of-plane deformations such as bending and twisting, whose curvatures are not always equal to zero. As indicated by the dotted black line in Figure 3, a bending mode occurs for $\kappa_{xy} = 0$; e.g., $[0°, 90°]$ and $[90°, 0°]$. The twisting mode appears with a non-zero twisting curvature $(\kappa_{xy} \neq 0)$, which is the most common deformation mode for bilayer angled plies.

The Gaussian curvature $K(= \kappa_1 \cdot \kappa_2)$ is often used to quantify a curved surface, where $\kappa_1$ and $\kappa_2$ are the principal curvatures obtained from $\kappa_x, \kappa_y$, and $\kappa_{xy}$ [33-36]. Notably, $K = 0$ and $K \neq 0$ for the in-plane and out-of-plane deformation modes, respectively. For layers of different thicknesses ($t_1 \neq t_2$), the bending regions are not mapped on a straight line but on a curved path, as shown in Figure 3b, caused by the asymmetric arrangement in the $z$-direction. However, in-plane deformations are not affected by the biased thickness arrangement. The deformation maps under other activation temperatures with various thickness ratio are shown in Figures S2-S4 in the Supplementary Information.

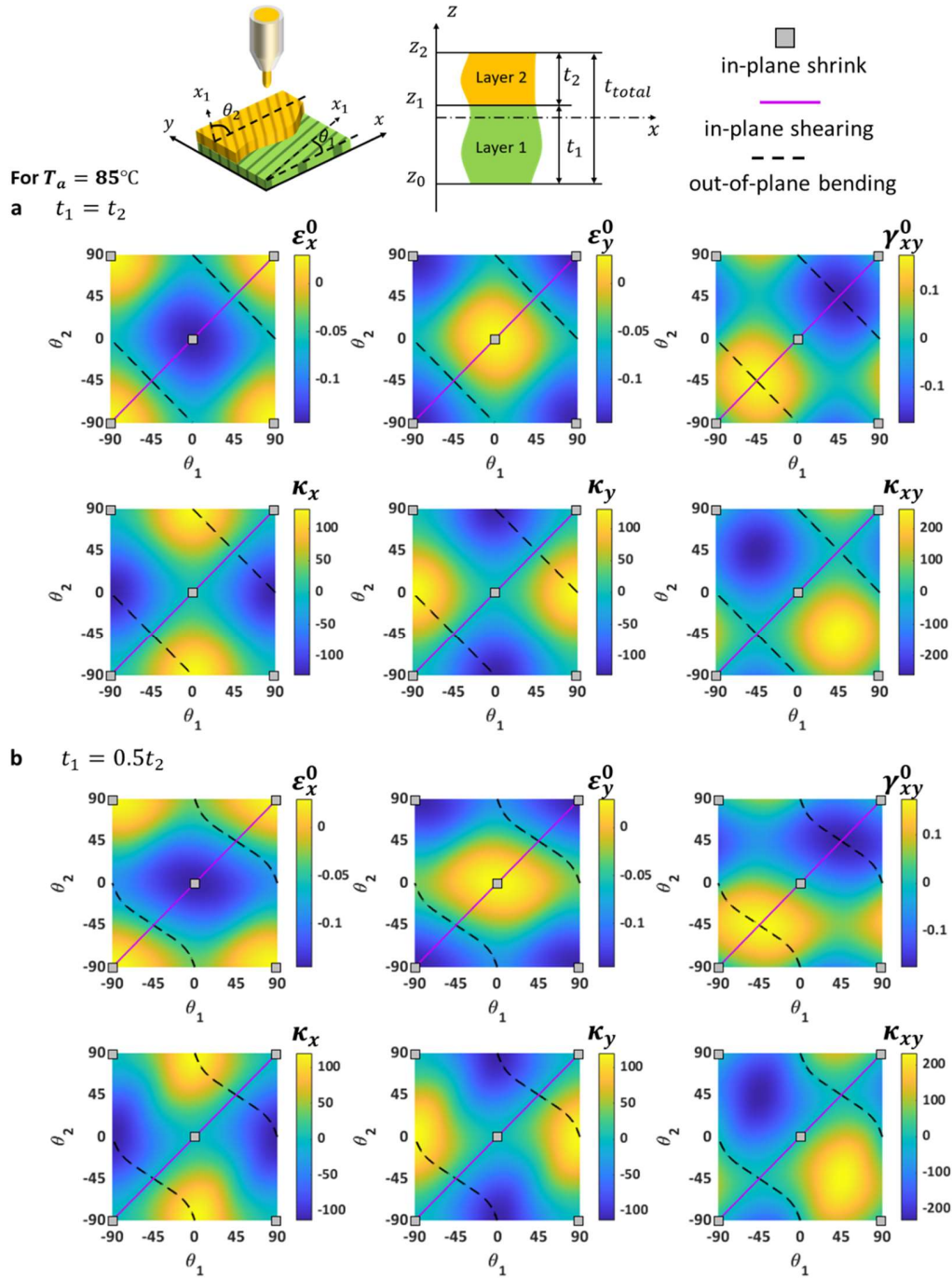

*Figure 3. Four modes of self-actuation of bilayer plates for varying printing parameters — printing angles and thickness ratio of bilayer at $T_a = 85°C$: (a) $t_1 = t_2$ and (b) $t_1 = 0.5t_2$; in-plane axial ($\varepsilon_x^0$ and $\varepsilon_y^0$) and shear ($\gamma_{xy}^0$) strains on the mid-plane and out-of-plane bending ($\kappa_x$ and $\kappa_y$) and twisting ($\kappa_{xy}$) curvatures.*

We can predict the deformation as a forward design from the three in-plane strains and three curvatures on the midplane. Notably, the plane curvatures $\kappa_x$ ($= -\partial^2 w/\partial x^2$) and $\kappa_y$ ($= -\partial^2 w/\partial y^2$) are not difficult to visualize, but the twisting curvature $\kappa_{xy}$ ($= -2\partial^2 w/\partial x \partial y$) is challenging to visualize [29, 30].

Therefore, we utilized two principal curvatures $\kappa_1$ and $\kappa_2$ from $\kappa_x$, $\kappa_y$, and $\kappa_{xy}$. Knowing that $\kappa_1 \cdot \kappa_2 < 0$ for a nonzero $\kappa_{xy}$, we can map the principal curvatures on the inner surface of a torus whose geometry can be featured by negative Gaussian curvature, whose geometric details are further described in Section IV.

Figure 4 demonstrates the thermomechanical deployment of printed bilayer plates with varying printing directions and layer thickness ratios. The individual deformations are indicated on the thermomechanical analysis map in Figure 4a. The analytical models and FE simulations show good agreement with the experiments for the four different transformation modes, as shown in Figure 4b. Most direct 4D printing methods show only a simple 1D deformation mode— bending [1-5]; even global complex deformation has been built with a lattice design by collecting gradients of 1D deformation of beams. In contrast, our approach can directly deploy a 2D plate to a 3D curved one. Moreover, we develop an inverse-design algorithm to provide processing parameters of a 2D plate for a given target 3D shape after transformation, as addressed in the next section.

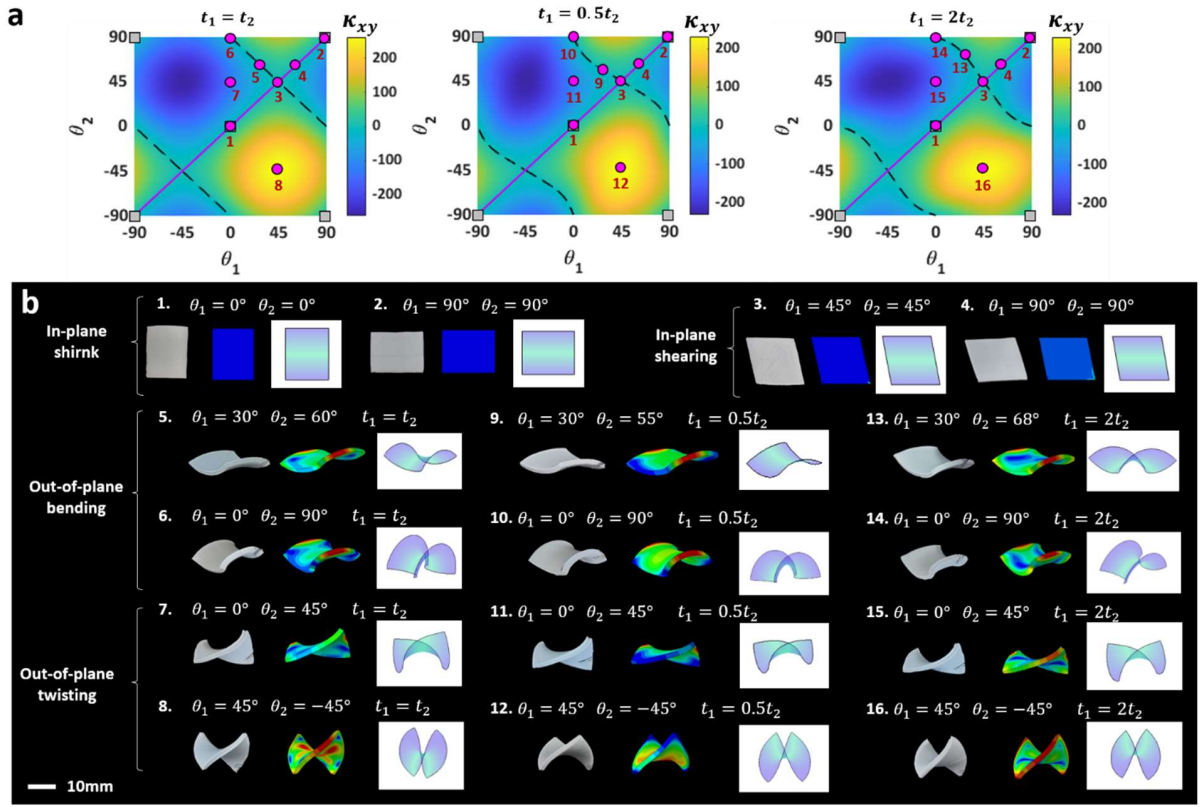

*Figure 4. Self-actuated shapes of printed bilayers with the four deployment modes at $T_a = 85°C$ for varying printing angles and thickness ratios; specific printing conditions were used for the demonstration: a nozzle traveling speed of $60\ mm/s$, a nozzle temperature of $210°C$, a layer height of $0.1\ mm$, and a flow compensation of $100\%$; The gray shapes are captured after experiment and the colored shapes are generated by the analytical model.*

## IV. Inverse design of doubly curved surfaces

For a given thickness ratio and $T_a$, we can deduce the dimensions and processing parameters of the initial plate — two lengths $a$ and $b$ and the printing path angles $\theta_1$ and $\theta_2$ for the target freeform plate in Figure 5a. The fundamental approach of the inverse design in this work is to determine the principal curvatures $\kappa_1$ and $\kappa_2$ and the transformation angle $\varphi$ from a target freeform surface and to find a transformation into the printing (structural) coordinate, $\kappa_x$, $\kappa_y$, and $\kappa_{xy}$.

Even though we can generate surfaces with positive Gaussian curvature, e.g., Figure 1d, we focus on doubly curved surfaces with a negative Gaussian curvature typically obtained by a bilayer printing in 2D cartesian coordinates. Torus is a representative geometry with a negative Gaussian curvature whose geometric features are determined by positive and negative curvatures at the inner hole circle and the tubular circle related to $r_1$ and $r_2$, respectively [37]. ~~To facilitate determination of the principal curvatures, we superimpose the freeform surface onto the surface of a torus; any freeform surface with a negative Gaussian curvature can map into the surface of a torus whose geometry can be featured with both positive and negative curvatures at the inner hole circle and the tubular circle featured by $r_1$ and $r_2$, respectively [37].~~ We can describe the Gaussian curvature of the surface of a torus as $K = \frac{\cos\beta}{r_2(R_h + r_2\cos\beta)}$, where $R_h$ is the distance between the centers of two circles, $O_1$ and $O_2$, expressed as $R_h = r_1 + r_2$, and $\beta$ is the central angle of the circle with $O_2$, as shown in Figure 5b. The Gaussian curvature can be decomposed into two principal curvatures $K(=(\kappa_1)_i \cdot \kappa_2)$ with $(\kappa_1)_i = \frac{\cos\beta}{R_h + r_2\cos\beta}$ and $\kappa_2 = \frac{1}{r_2}$.

$(\kappa_1)_i$ in a torus varies along the $z$-axis with a maximum value along a central line of the inner hole's circle, as indicated in Figure 5b. However, $\kappa_2$ in a torus does not change at any points on the radius of the tubular circle. We describe the details of the derivation of $(\kappa_1)_i$ and $\kappa_2$ from a freeform surface in Sections III and IV of the Supplementary Information. The positions having the maximum $(\kappa_1)_i$ in a torus are located on the inner middle circle at $r = r_1$ in Figure 5b; $(\kappa_1)_{max} = 1/r_1$. We plot the curvatures of a freeform surface in Figure 5c whose positions at the corners are $(x_i, y_i, z_i)$ with $i = A, B, C, D$. By aligning $\kappa_1$ of the freeform surface with $(\kappa_1)_{max}$ in a torus, as illustrated by the dashed line in Figure 5d, positions $A$ and $C$ have the same $\kappa_1$ yet $z_A = -z_C$. Similarly, positions $B$ and $D$ have the same $\kappa_1$ and $z_B = -z_D$. Using these relations, a 2D plot of the stretched freeform surface is presented in Figure 5e, indicating the relationship of the principal and printing coordinates with the transformation angle $\varphi$, as shown in Figures 5h and 5j. Notably, the edges of the outer rectangle in Figure 5e are parallel to the principal curvature coordinate, while $AB$ of the interior rectangle represents the $x$-axis of the printing coordinate in Figure 5e. Notably, the length and width of the inner rectangle in purple are the arc lengths of the inner middle circle and tubular circle of the torus surface where the freeform surface is superimposed, respectively.

From the projection of the freeform surface to the top view and the tubular cross-section, we can obtain the lengths of $BM$ and $AM$, as indicated by the red lines in Figures 5f and 5g. Once we determine $AB$ and $AM$, we can obtain $\varphi$ $(= \tan^{-1} \overline{BM}/\overline{AM})$ from the geometric relationship in Figure 5e. We describe the details of the derivation of $\varphi$ in Section IV of the Supplementary Information.

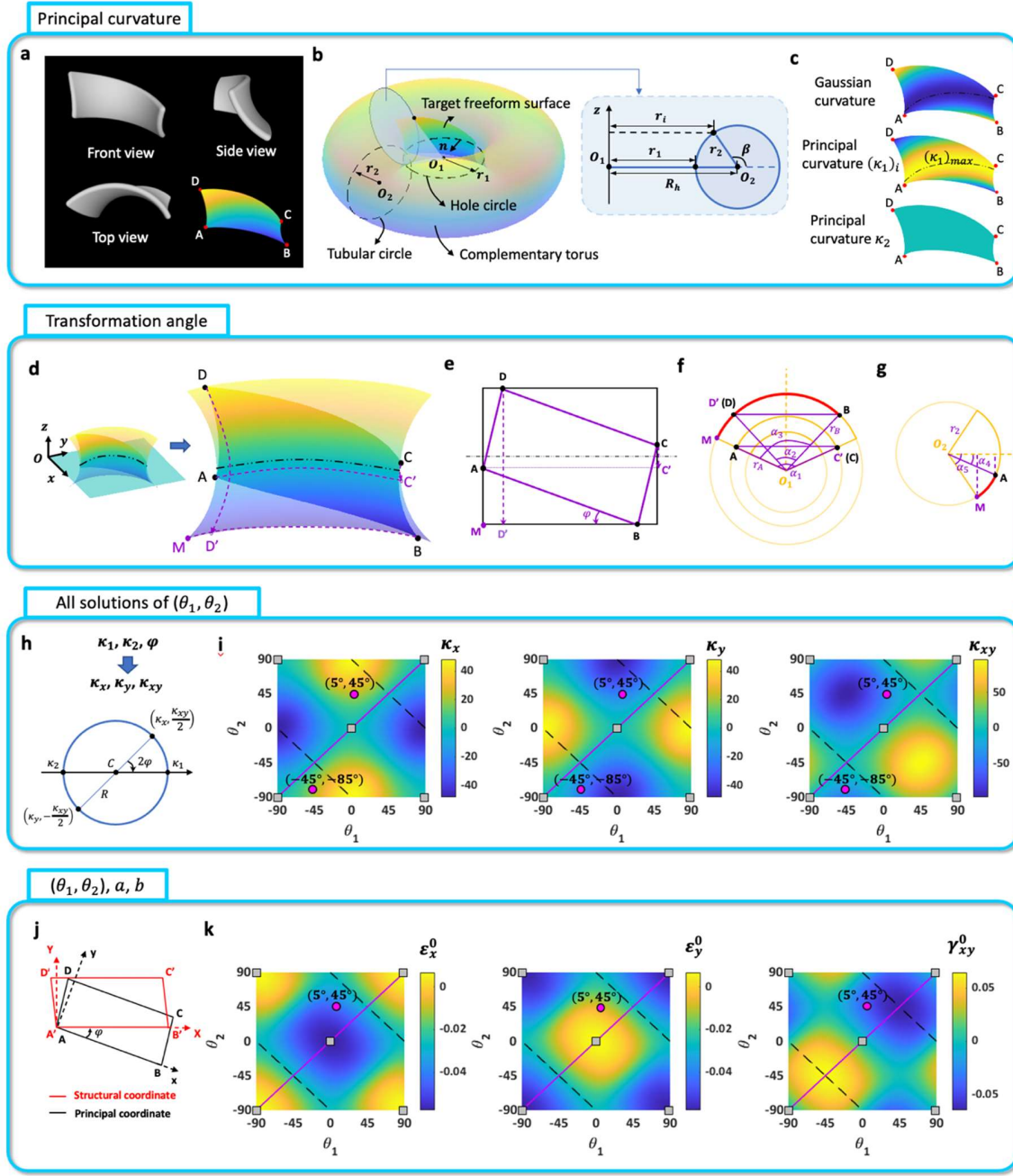

*Figure 5. Procedure of an inverse design of a freeform surface: (a) 3D target shape with four corner points A, B, C, and D; (b) the target shape mapped onto a torus defined by hole ($r_1$) and tubular ($r_2$) circles with an Illustration of the cross-section of a torus, showing the radii of the circles at the four corners $r_i$; (c) Gaussian curvature K with two principal curvatures $\kappa_1$ and $\kappa_2$ of the target shape onto a torus's inner surface; (d) the target shape superimposed on the inner torus surface while aligning $\kappa_1$ on the xy plane, where the black dashed line indicates $(\kappa_1)_{max}$; (e) a 2D flattened shape of the target geometry superimposed on an inner torus surface; (f) top view of the torus showing a series of concentric circles centered at $O_1$ with various radii; (g) cross-sectional view of the torus showing the tubular circle centered at $O_2$; (h) transformation from ($\kappa_1$, $\kappa_2$, $\varphi$) to ($\kappa_x$, $\kappa_y$, $\kappa_{xy}$) using Mohr's circle, (i) all possible ($\theta_1$, $\theta_2$) satisfying the desired curvatures from the design maps; (j) transformation of the principal curvature coordinates ($x_i$, $y_i$) into the structural coordinates ($X_i$, $Y_i$); (k) filtering procedure to finalize ($\theta_1$, $\theta_2$).*

Once we know $\kappa_1$, $\kappa_2$, and $\varphi$, we can obtain $\kappa_x$, $\kappa_y$, and $\kappa_{xy}$ using Mohr's circle of curvature, as shown in Figure 5h. Subsequently, knowing $\kappa_x$, $\kappa_y$, and $\kappa_{xy}$ provides the printing paths $\theta_1$ and $\theta_2$ from the thermomechanical analysis map in Figure 5i. Notably, knowing $\theta_1$ and $\theta_2$ also provides us with the in-plane axial and shear strains: $\varepsilon_x^0, \varepsilon_y^0$, and $\gamma_{xy}^0$ in Figure 5k. Once the in-plane strains are known, we can obtain the dimensions of the initial rectangular plate $a$ and $b$ using the deformation gradient $\mathbf{F}$, where we describe the details of the derivation in Section V of the Supplementary Information.

Previously, groups have also developed an inverse-design algorithm of direct 4D printing [1,2]; their approach is to find the processing parameters for a collection of 1D beams made of multi-materials for a target 3D shape. However, our inverse-design goal is to find the processing parameters of a 2D plate made of a single material for a target 3D surface.

Notably, our analytical model with a bilayer plate can be easily extended to multi-layer plates using the laminated plate theory. However, the deformation modes of the multi-layer plates still belong to the four modes of this study. Deployment of multi-layer plates only provides different magnitudes of in-plane strain and curvature.

## V. Multi-modal direct 4D printing of complex 3D shapes

Using the thermomechanical deformation model of a bilayer plate integrated with the anisotropic shape memory effect and curvature of the target surface during the 3D printing of a single isotropic material, we can build complex structures directly deployed from a flat surface. Figures 6a and 6b show a square lattice that can directly deploy chiral and achiral lattices with the bending mode of the members (Supplementary Video 2). Figure 6c demonstrates the significant deformation of a bilayer flat beam transforming into a lattice sphere (Supplementary Video 3). Note that conventional thermal expansion mismatch could not deploy such a large deformation [26,38]; only the shape memory effect can produce large deformations. In addition, note that all structures can remain stiff after deployment due to the shape fixity of printing material at room temperature. Figure 6d shows the transformation of a flat flower with a bending and a twisting modal deployment for pedicels and petals, respectively (Supplementary Video 4). Note that sequential deformation is possible by varying the thickness of the pedicels [15]. Figures 6e–6g demonstrate a direct 4D printing of complex animal shapes with four modes — in-plane axial and shear and out-of-plane bending and twisting (Supplementary Video 5).

Previously, one group demonstrated individual control of multiple units of three independent components with two materials, and specially designed bending joints produced global curvature [1]. The method is complicated to implement with high-resolution printing of each component with multi-materials. Moreover, it produces local stress by contact with adjacent units even after deployment. However, our method can produce stress-free complex shapes with four individual modes only using a single material printing, which is significant because of its simple implementation and potential for broad applications, considering inexpensive and ubiquitous 3D printing machines such as FDM and DIW, whose market shares are already ~77% in the world [39].

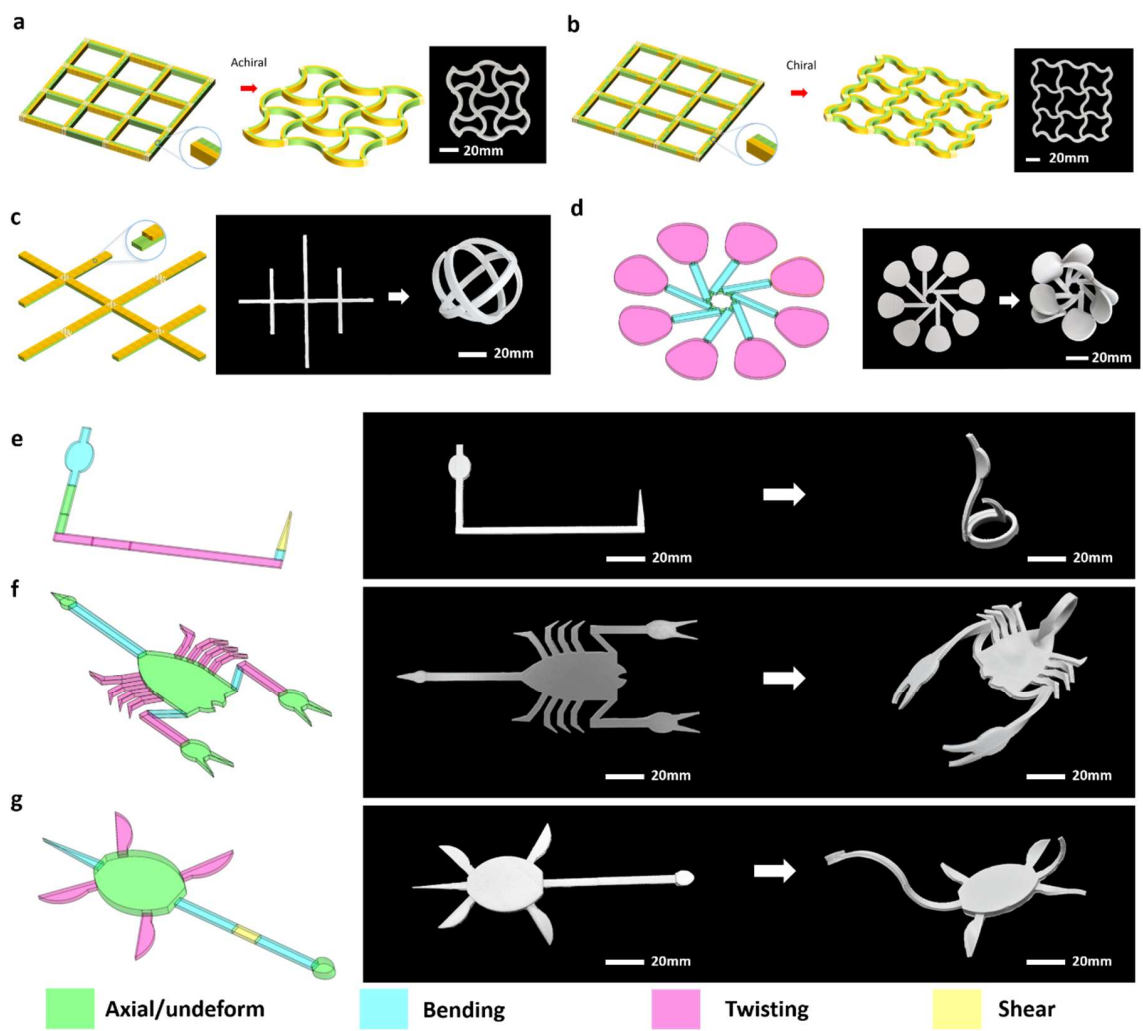

*Figure 6. Examples of single and multi-modal deployments of the direct 4D printing of a single material: (a) chiral and (b) achiral square lattices with a bending mode; (c) sphere lattice with a bending mode; (d) bimodal transformation of a flower, petal with bending and twisting modes; and (e) snake, (f) scorpion, and (g) plesiosaur with multi-modal deformations, in-plane axial and shear and out-of-plane bending and twisting. Video files are available in the Supplementary Information.*

## VI. Discussion

Morphing flat sheets into complex 3D shapes is extremely useful for fast manufacturing and saving materials without supporting materials, while they also have volumetrically efficient storage and shipment and functional usage. Direct 4D printing has the potential to morph complex 3D shapes out of the as-printed 2D plates. Most direct 4D printing use 1D structural deformation — bending and coiling to morph a flat shape by controlling only a single curvature along the longitudinal direction. However, many surfaces are continuous 3D curved ones with nonzero Gaussian curvature in nature. We demonstrate that the extrusion-based 3D printing of an isotropic single material system at an elevated temperature can directly morph an as-printed sheet into a doubly-curved surface.

Integration of the intrinsic thermomechanical training of an SMP during extrusion-based printing with the laminated plate theory enables us to construct an inverse design method for a direct 4D printing of doubly-curved surfaces. The mapping of target surfaces with negative Gaussian curvature onto the inner surface of a torus is a powerful tool for direct 4D printing. However, our current model is built based on a unit cell level. For more complex geometries with multiple Gaussian curvatures, one may consider the discretization of various units and connection among adjacent units. Moreover, the current model cannot generate a surface with positive Gaussian curvature such as a sphere and pseudosphere, which can be future work. Despite the limitation, our method can produce stress-free complex shapes with multi modes and shape locking using only a single material printing, which is significant because of its simple implementation and potential for broad applications such as biomedical devices, deployable structures, intelligent textiles, soft robots, active metamaterials, and pop-up Kirigami structures

## VII. Materials and Methods

**Fabrication and morphing of bilayer plates**

We printed flat bilayer plates with a polylactic acid (**PLA**) filament (PolyLite PLA, Polymaker) using an FDM-based 3D printer (Ultimaker 2+, Ultimaker, The Netherlands). Using a commercial preprocessing code (Ultimaker Cura 4.8.0, Ultimaker, The Netherlands) of FDM, we sliced the plate with two layers. We implemented the bilayer design with varying layer thicknesses and printing angles using the Support Blocker function of Cura. Other important printing parameters included the nozzle temperature and the printing-layer height. After fabrication, we deployed the bilayer plate in a hot water tank ($T_a \geq 65°C$), whose temperature was controlled by an immersion heater (Anova, USA).

**Material characterization**

We printed single-layer PLA strips ($30 \times 4 \times 1\ mm^3$) with a 0° printing angle. Using **DMA** (DMA Q850, TA Instruments, DE, USA), we measured the storage modulus and $T_g$ of PLA. We varied temperature from $25°C$ to $85°C$ with a temperature ramping option at $3°C/min$ for a tensile loading mode with a frequency of 1 Hz, as shown in Figure 2a. Following van Manen et al. [15], we printed single-layer PLA strips ($30 \times 10 \times 1 mm^3$) with a 0° printing angle. We measured the thermomechanical shrinkage of the PLA strips using an image processing tool of MATLAB (R2020a, MathWorks, USA) and determined the activation time at which the deformation reached a stable state for varying sets of temperature exposure in hot water ($\geq 65°C$) for $180\ s$. We also characterized the mechanical properties of a printed single-layer PLA having a transverse isotropic property. The detailed procedure to obtain the parameters is shown in Section II of the Supplementary Information.

**Finite-element simulation**

We simulated the thermomechanical deformation of the bilayer plates during recovery of the SMP using the commercial finite-element (FE) code ABAQUS (Dassault Systems, SIMULIA Corp., USA). We constructed a square (or rectangular) 3D deformable shell model with a 4-node linear shell element (S4R in ABAQUS) for varying thickness ratios and printing path angles of each lamina of the bilayer plate. To simulate the expansion or compression of each layer in the orthotropic directions during the deployment stage, we used a fictitious thermal expansion [40] of the elastic material and applied a temperature change in a predefined field. For each specific $T_a$, the fictitious coefficients of thermal expansion in two directions of a unidirectional laminate were defined as $\alpha_1^T = \varepsilon_1^T/\Delta T$ and $\alpha_2^T = \varepsilon_2^T/\Delta T$, where $\Delta T$ is the change in temperature, and $\varepsilon_1^T$ and $\varepsilon_2^T$ are the thermal strains of the unidirectional lamina from the experiment.

## VIII. Conclusion

In summary, we here encoded a direct 4D printing of a single-material system that can deploy with mixed multimodal and doubly curved shapes — in-plane extension and shear and out-of-plane bending and twisting from as-printed plate shapes. Using the intrinsic anisotropic shape memory effect during the extrusion-based 3D printing process, we quantitatively predicted the 3D shapes after deployment from the printed 2D geometries by considering the processing parameters such as the printing angles and bilayer thickness ratio, printing temperature, nozzle speeds. Our inverse-design algorithm considering both in-plane and out-of-plane deployments suggests the printing and deployment conditions of a 2D bilayer plate for a targeted complex 3D shape, including doubly curved shapes. In addition to the advantages of direct 4D printing, such as fast printing time with 2D geometries and no need for supporting materials, our 4D printing approach with a single material can easily be implemented with inexpensive and ubiquitous 3D printing machines such as FDM and DIW, whose market shares are already ~77% in the world. Because of its shape-locking and stress-free features after morphing, our method is uniquely advantageous for various structural applications in the robotic, automotive, aerospace, and biomedical engineering fields such as pop-up kirigami structures, intelligent costumes, autonomous robotics, vehicle components, biomedical devices, and tissue engineering. This approach opens a new avenue of direct 4D printing of continuous bilayer plates with a single-material system through an inverse-design algorithm coupling materials, processes, deployment, and structural mechanics.

## Acknowledgements

The authors acknowledge support received from the Shanghai NSF (Award # 17ZR1414700) and the Research Incentive Program of Recruited Non-Chinese Foreign Faculty by Shanghai Jiao Tong University. We thank Prof. Xiang Zhou at the School of Aeronautic and Astronautic Engineering in Shanghai Jiao Tong University for assistance with the mechanical testing of the materials.

## Author contributions

J.J. designed and supervised the research; B.Z. and C.S. built the analytical models; C.S. designed the deformed structures and performed the experiments; Z.H. performed the mechanical characterization of the materials; B.Z. performed the coding, FE simulation, and experiments; B.Z. and Z.H. conceived the inverse-design algorithm; B.Z., C.S, and J.J. wrote the manuscript and prepared the figures.

## Supplementary information

Supporting Information is available.

## Data availability

MATLAB codes are available on GitHub. link